# Kinetics of a mixed spin-1/2 and spin-3/2 Ising ferrimagnetic model

**Bayram Deviren[a], Mustafa Keskin[b, *], Osman Canko[b]**

[a] *Institute of Science, Erciyes University, 38039 Kayseri, Turkey*
[b] *Department of Physics, Erciyes University, 38039 Kayseri, Turkey*

We present a study, within a mean-field approach, of the kinetics of a mixed ferrimagnetic model on a square lattice in which two interpenetrating square sublattices have spins that can take two values, $\sigma = \pm 1/2$, alternated with spins that can take the four values, $S = \pm 3/2, \pm 1/2$. We use the Glauber-type stochastic dynamics to describe the time evolution of the system with a crystal-field interaction in the presence of a time-dependent oscillating external magnetic field. The nature (continuous and discontinuous) of transition is characterized by studying the thermal behaviors of average order parameters in a period. The dynamic phase transition points are obtained and the phase diagrams are presented in the reduced magnetic field amplitude (h) and reduced temperature (T) plane, and in the reduced temperature and interaction parameter planes, namely in the (h, T) and (d, T) planes, d is the reduced crystal-field interaction. The phase diagrams always exhibit a tricritical point in (h, T) plane, but do not exhibit in the (d, T) plane for low values of h. The dynamic multicritical point or dynamic critical end point exist in the (d, T) plane for low values of h. Moreover, phase diagrams contain paramagnetic (p), ferromagnetic (f), ferrimagnetic (i) phases, two coexistence or mixed phase regions, (f+p) and (i+p), that strongly depend on interaction parameters.

## 1. Introduction

In last two decades, mixed spin Ising systems have attracted a great deal of attention. The reasons are follows: (i) These problems are mainly related to the potential technological applications in the area of thermomagnetic recording [1]. (ii) The systems have less translational symmetry than their single spin counterparts; hence exhibit many new phenomena that cannot be observed in the single-spin Ising systems. (iii) The study of these systems can be relevant for understanding of bimetallic molecular systems based magnetic materials [2]. One of the well known mixed spin Ising systems is the mixed spin-1/2 and spin-3/2 Ising model. Amorphous V(TCNE)$_x$. $y$ (solvent), where TCNE is tetracyanoethylene, are organometallic compounds that seem to have a 1/2 - 3/2 ferrimagnetic structure and order ferrimagnetically as high as 400K [3, 4].

An early attempt to study the magnetic properties of the diluted mixed spin-1/2 and spin-3/2 Ising model Hamiltonian with only a bilinear exchange interaction (J) was made with in the framework of the effective-field theory (EFT) by Bobák and Jurčišin [5]. They found that the compensation point which depends not only on the magnitude of spins but also on the lattice structure. Bobák and Jurčišin [6] investigated the diluted mixed spin-1/2 and spin-3/2 Ising model Hamiltonian with J and the crystal-field (D) interactions on the honeycomb lattice within the EFT and found that the system exhibit two compensation points. Benayad et al. [7] studied the mixed spin-1/2 and spin-3/2 Ising model Hamiltonian with J and the crystal-field (D) interactions on the honeycomb lattice by using the EFT, and they found a variety of interesting phenomena in phase diagrams due to the influence of the crystal-field interaction. Magnetic properties of the mixed spin-1/2 and spin-3/2 transverse Ising model with a crystal-field interaction were studied within the EFT, extensively [8]. Especially, the thermal behavior of order parameters are investigated and phase diagrams are presented. Monte Carlo (MC) study of a mixed spin-1/2 and spin-3/2 Ising model on a square lattice was done by Buendia and Cardona [9], and observed that the compensation temperatures are extremely dependent on the interactions in the Hamiltonian. Magnetic properties of the mixed spin-1/2 and spin-3/2 Ising model in a longitudinal magnetic field were investigated, and thermal behaviors of magnetizations, magnetic susceptibilities and the phase diagram are examined in detail [10]. Li et al. [11] studied the mixed spin-1/2 and spin-3/2 quantum Heisenberg system on a square lattice with the double-time-temperature Green function method to investigate the effects of the nearest- and next- nearest-neighbor interactions between spins on the magnetic behavior of the system, especially on the compensation point. The system has also been investigated on the Bethe lattice [12] and two-fold Cayley tree [13] using the exact recursion relations, on the honeycomb lattice within the framework of an exact star-triangle mapping transformations [14], and on the extended Kagomé lattice [15] and union Jack (centered square) lattice [16] by establishing a mapping correspondence with the eight-vertex model.

---

* Corresponding author.
Tel: + 90 (352) 4374938#33105; Fax: + 90 (352) 4374931
E-mail address: keskin@erciyes.edu.tr (M. Keskin)

Despite of all these equilibrium studies, as far as we know, the nonequilibrium aspects of this system have not been investigated. Therefore, the purpose of the present work is to investigate dynamical aspect of the mixed spin-1/2 and spin-3/2 Ising ferrimagnetic model with a crystal-field interaction in the presence of a time-dependent oscillating external magnetic field. We use the Glauber-type stochastic dynamics [17] to describe the time evolution of the system. The nature (continuous and discontinuous) of transition is characterized by studying the thermal behaviors of average order parameters in a period. The dynamic phase transition (DPT) points are obtained and the dynamic phase diagrams are presented in different planes.

The organization of the remaining part of this paper is as follows. In Section 2, the model and its formulations, namely the derivation of the set of mean-field dynamic equations, are given by using Glauber-type stochastic dynamics in the presence of a time-dependent oscillating external magnetic field. In Section 3, we solve the coupled set of dynamic equations and present the behaviors of time variations of order parameters and the behavior of the average order parameters in a period, which are also called the dynamic order parameters, as functions of the reduced temperature and as a result, the DPT points are calculated. Section 4 contains the presentation and the discussion of the dynamic phase diagrams. Finally, summary and conclusion are given in Section 5.

## 2. Model and formulations

The mixed spin-1/2 and spin-3/2 Ising model is described as a two-sublattice system, with spin variables $\sigma_i = \pm 1/2$ and $S_j = \pm 3/2, \pm 1/2$ on the sites of sublattices A and B, respectively. The system has two long-range order parameters, namely the average magnetizations $<\sigma>$ and $<S>$ for the A and B sublattices, respectively, which are the excess of one orientation over the other, also called the dipole moments.

The Hamiltonian of the mixed spin-1/2 and spin-3/2 Ising model with the bilinear (*J*) nearest-neighbor pair interaction and a single-ion potential or crystal-field interaction (*D*) in the presence of a time-dependent oscillating external magnetic field is

$$\mathcal{H} = -J\sum_{\langle ij \rangle} \sigma_i^A S_j^B - D\sum_j \left[(S_j^B)^2 - 5/4\right] - H\left(\sum_i \sigma_i^A + \sum_j S_j^B\right), \quad (1)$$

where <*ij*> indicates a summation over all pairs of nearest-neighboring sites, and *H* is an oscillating magnetic field of the form

$$H(t) = H_0 \cos(wt), \quad (2)$$

where $H_0$ and $w = 2\pi v$ are the amplitude and the angular frequency of the oscillating field, respectively. The system is in contact with an isothermal heat bath at absolute temperature.

Now, we apply Glauber-type stochastic dynamics [17] to obtain the mean-field dynamic equation of motion. Thus, the system evolves according to a Glauber-type stochastic process at a rate of $1/\tau$ transitions per unit time. Leaving the *S* spins fixed, we define $P^A(\sigma_1, \sigma_2, \ldots, \sigma_N; t)$ as the probability that the system has the σ-spin configuration, $\sigma_1, \sigma_2, \ldots, \sigma_N$, at time t, also, by leaving the σ spins fixed, we define $P^B(S_1, S_2, \ldots, S_N; t)$ as the probability that the system has the S-spin configuration, $S_1, S_2, \ldots, S_N$, at time t. Then, we calculate $W_i^A(\sigma_i)$ and $W_j^B(S_j \to S_j')$, the probabilities per unit time that the *i*th σ spin changes from $\sigma_i$ to $-\sigma_i$ (while the spins on B sublattice momentarily fixed) and the *j*th *S* spin changes from $S_j$ to $S_j'$ (while the spins on A sublattice momentarily fixed), respectively. Thus, if the spins on the sublattice B momentarily fixed, the master equation for the sublattice A can be written as

$$\frac{d}{dt} P^A(\sigma_1, \sigma_2, \ldots, \sigma_N; t) = -\sum_i W_i^A(-\sigma_i) P^A(\sigma_1, \sigma_2, \ldots, \sigma_i, \ldots \sigma_N; t) \\ + \sum_i W_i^A(\sigma_i) P^A(\sigma_1, \sigma_2, \ldots, -\sigma_i, \ldots \sigma_N; t). \quad (3)$$

Since the system is in contact with a heat bath at absolute temperature $T_A$, each spin σ can flip with the probability per unit time;

$$W_i^A(\sigma_i) = \frac{1}{\tau} \frac{\exp(-\beta \Delta E^A(\sigma_i))}{\sum_{\sigma_i} \exp(-\beta \Delta E^A(\sigma_i))}, \quad (4)$$

where $\beta = 1/k_B T_A$, $k_B$ is the Boltzmann factor, $\sum_{\sigma_i}$ is the sum over the two possible values of $\sigma_i^A$, $\pm 1/2$, and

$$\Delta E^A(\sigma_i) = 2\sigma_i(H + J\sum_j S_j), \qquad (5)$$

gives the change in the energy of the system when the $\sigma_i$-spin changes. The probabilities satisfy the detailed balance condition

$$\frac{W_i^A(-\sigma_i)}{W_i^A(\sigma_i)} = \frac{P^A(\sigma_1, \sigma_2, \ldots, -\sigma_i, \ldots \sigma_N)}{P^A(\sigma_1, \sigma_2, \ldots, \sigma_i, \ldots \sigma_N)}, \qquad (6)$$

and substituting the possible values of $\sigma_i$, we get

$$W_i^A(-\frac{1}{2}) = \frac{1}{2\tau} \frac{\exp(-\beta x/2)}{\cosh(\beta x/2)}, \qquad (7a)$$

$$W_i^A(\frac{1}{2}) = \frac{1}{2\tau} \frac{\exp(\beta x/2)}{\cosh(\beta x/2)}, \qquad (7b)$$

where $x = H + J\sum_j S_j$. From the master equation associated with the stochastic process, it follows that the average $<\sigma_k>$ satisfies the following equation [18]

$$\tau \frac{d}{dt}\langle\sigma_k\rangle = -\langle\sigma_k\rangle + \frac{1}{2}\tanh\left[\frac{\beta}{2}\left(H + J\sum_j S_j\right)\right]. \qquad (8)$$

This dynamic equation can be written in terms of a mean-field approach and hence the first mean-field dynamical equation of the system in the presence of a time-varying field is:

$$\Omega \frac{d}{d\xi} m_A = -m_A + \frac{1}{2}\tanh\left[\frac{1}{2T}(m_B + h\cos(\xi))\right], \qquad (9)$$

where $m_A = \langle\sigma\rangle$, $m_B = \langle S\rangle$, $\xi = wt$, $T = (\beta zJ)^{-1}$, $h = H_0/zJ$ and $\Omega = \tau w$.

Now assuming that the spins on sublattice A remain momentarily fixed and the spins on the sublattice B change, we obtain the mean-field dynamical equation of $m_B$ for the B sublattice. Since $S_j = \pm 3/2, \pm 1/2$, the master equation for the sublattice B can be written as

$$\frac{d}{dt}P^B(S_1,S_2,...,S_N;t) = -\sum_j\left(\sum_{S_j\neq S_j'}W_j^B(S_j\to S_j')\right)P^B(S_1,S_2,...,S_j,...,S_N;t)$$
$$+\sum_j\left(\sum_{S_j\neq S_j'}W_j^B(S_j'\to S_j)P^B(S_1,S_2,...,S_j',...,S_N;t)\right), \quad (10)$$

where $W_j^B(S_j\to S_j')$ is the probability per unit time that the *j*th spin changes from the value $S_j$ to $S_j'$, and in this sense the Glauber model is stochastic. Since the system is in contact with a heat bath at absolute temperature $T_A$, each spin can change from the value $S_j$ to $S_j'$ with the probability per unit time;

$$W_j^B(S_j\to S_j') = \frac{1}{\tau}\frac{\exp(-\beta\Delta E^B(S_j\to S_j'))}{\sum_{S_j'}\exp(-\beta\Delta E^B(S_j\to S_j'))}, \quad (11)$$

where $\sum_{S_j'}$ is the sum over the four possible values of $S_j'$, $\pm 3/2, \pm 1/2$, and

$$\Delta E^B(S_j\to S_j') = -(S_j'-S_j)(H+J\sum_i\sigma_i) - \left[(S_j')^2-(S_j)^2\right]D, \quad (12)$$

gives the change in the energy of the system when the $S_j$-spin changes. Using the detailed balance condition and substituting the possible values of $S_j$, we get

$$W_j^B(\frac{3}{2}\to -\frac{3}{2}) = W_j^B(\frac{1}{2}\to -\frac{3}{2}) = W_j^B(-\frac{1}{2}\to -\frac{3}{2})$$
$$= \frac{1}{2\tau}\frac{\exp(-\beta D)\exp(-3\beta y/2)}{\exp(\beta D)\cosh(3\beta y/2)+\exp(-\beta D)\cosh(\beta y/2)}, \quad (13a)$$

$$W_j^B(\frac{3}{2}\to -\frac{1}{2}) = W_j^B(\frac{1}{2}\to -\frac{1}{2}) = W_j^B(-\frac{3}{2}\to -\frac{1}{2})$$
$$= \frac{1}{2\tau}\frac{\exp(-\beta D)\exp(-\beta y/2)}{\exp(\beta D)\cosh(3\beta y/2)+\exp(-\beta D)\cosh(\beta y/2)}, \quad (13b)$$

$$W_j^B(\frac{3}{2}\to \frac{1}{2}) = W_j^B(-\frac{1}{2}\to \frac{1}{2}) = W_j^B(-\frac{3}{2}\to \frac{1}{2})$$
$$= \frac{1}{2\tau}\frac{\exp(-\beta D)\exp(\beta y/2)}{\exp(\beta D)\cosh(3\beta y/2)+\exp(-\beta D)\cosh(\beta y/2)}, \quad (13c)$$

$$W_j^B(\tfrac{1}{2}\to\tfrac{3}{2})=W_j^B(-\tfrac{1}{2}\to\tfrac{3}{2})=W_j^B(-\tfrac{3}{2}\to\tfrac{3}{2})$$
$$=\frac{1}{2\tau}\frac{\exp(\beta D)\exp(3\beta y/2)}{\exp(\beta D)\cosh(3\beta y/2)+\exp(-\beta D)\cosh(\beta y/2)},\quad (13d)$$

where $y=H+J\sum_i \sigma_i$. Notice that, since $W_j^B(S_j\to S_j')$ does not depend on the value $S_j$. We can therefore write $W_j^B(S_j\to S_j')=W_j^B(S_j')$, then the master equation becomes

$$\frac{d}{dt}P^B(S_1,S_2,\ldots,S_N;t)=-\sum_j\left(\sum_{S_j\neq S_j'}W_j^B(S_j')\right)P^B(S_1,S_2,\ldots,S_j,\ldots,S_N;t)$$
$$+\sum_j W_j^B(S_j)\left(\sum_{S_j\neq S_j'}P^B(S_1,S_2,\ldots,S_j',\ldots,S_N;t)\right),\quad (14)$$

Since the sum of probabilities is normalized to one, by multiplying both sides of Eq. (14) by $S_j$ for $m_B$ and taking the average, we obtain

$$\tau\frac{d}{dt}\langle S_j\rangle=-\langle S_j\rangle+\left\langle\frac{3\exp(\beta D)\sinh(3\beta y/2)+\exp(-\beta D)\sinh(\beta y/2)}{2\exp(\beta D)\cosh(3\beta y/2)+2\exp(-\beta D)\cosh(\beta y/2)}\right\rangle,\quad (15)$$

This dynamic equation can be written in terms of a mean-field approach; hence the second mean-field dynamical equation of the system in the presence of a time-varying field is:

$$\Omega\frac{d}{d\xi}m_B=-m_B$$
$$+\frac{3\exp(d/T)\sinh[3(m_A+h\cos\xi)/2T]+\exp(-d/T)\sinh[(m_A+h\cos\xi)/2T]}{2\exp(d/T)\cosh[3(m_A+h\cos\xi)/2T]+2\exp(-d/T)\cosh[(m_A+h\cos\xi)/2T]},\quad (16)$$

where $d=D/zJ$. Thus, the set of the mean-field dynamical equations for the average magnetizations are obtained, namely Eqs. (9) and (16). We fixed $z=4$ and $\Omega=2\pi$. In the next section, we will give the solution and discussions of the set of coupled mean-field dynamical equations.

**3. Thermal behaviors of dynamic order parameters and dynamic phase transition**

In this section, we first investigate the behaviors of time variations of magnetizations and then the thermal variation of the average magnetizations in a period, which are also called the dynamic magnetizations, as functions of the reduced temperature and as a result the nature of transition is found and the DPT points are calculated. We also investigate the behavior of the dynamic magnetizations as a function of the reduced crystal-field interaction. For these purposes, first we have to study the stationary solutions of the set of coupled mean-field dynamical equations, given in Eqs. (9) and (16), when the parameters T, d and h are varied. The stationary solutions of these equations will be periodic functions of ξ with period 2π; that is, $m_A(\xi+2\pi)=m_A(\xi)$ and $m_B(\xi+2\pi)=m_B(\xi)$. Moreover, they can be one of third types according to whether they have or do not have the property

$$m_A(\xi+\pi)=-m_A(\xi) \quad \text{and} \quad m_B(\xi+\pi)=-m_B(\xi). \tag{17}$$

The first type of solution satisfies both Eq. (17) is called a symmetric solution which corresponds to a paramagnetic (p) solution. In this solution, the submagnetizations $m_A$ and $m_B$ are equal to each other ($m_A=m_B$) and $m_A(\xi)$ and $m_B(\xi)$ oscillate around zero and are delayed with respect to the external magnetic field. The second type of solution which does not satisfy Eq. (17), is called a nonsymmetric solution that corresponds to a ferromagnetic solution. In this solution, the submagnetizations $m_A$ and $m_B$ are equal each other ($m_A=m_B$). In this case the magnetizations do not follow the external magnetic field any more, but instead of oscillating around zero; they oscillate around a nonzero value, namely ±1/2; hence, we have the ferromagnetic ±1/2 (f) phase. The third type of solution, which does not satisfy Eq. (17), is also called a nonsymmetric solution but this solution corresponds to a ferrimagnetic (i) solution because the submagnetizations $m_A$ and $m_B$ are not equal to each other, and $m_A(\xi)$ and $m_B(\xi)$ oscillate around ±1/2 and ±3/2, respectively. These facts are seen explicitly by solving Eqs. (9) and (16) within the Adams-Moulton predictor-corrector method for a given set of parameters and initial values and presented in Fig. 1. From Fig. 1, one can see following five different solutions or phases, namely the p, f and i fundamental phases or solutions, and two coexistence phases or solutions, namely the f + p in which f and p solutions coexist; the i + p in which i and p solutions coexist, have been found. In Fig. 1(a) only the symmetric solution is always obtained, in this case $m_A=m_B$ oscillate around zero value $(m_A(\xi)=m_B(\xi)=0)$. Hence, we have a paramagnetic (p) solution or phase. On the other hand in Fig. 1 (b) and (c) only the nonsymmetric solutions are found; therefore, we have the f and i solutions, respectively. In Fig. 1(b), $m_A(\xi)$ and $m_B(\xi)$ oscillate around ±1/2; hence we have the ferromagnetic ±1/2 (f) phase. In Fig. 1(c), $m_A(\xi)$ oscillates around ±1/2 and $m_B(\xi)$ oscillates around ±3/2, this solution corresponds to the ferrimagnetic (i) phase $(m_A(\xi)\neq m_B(\xi)\neq 0)$. In Fig. 1(d), $m_A(\xi)$ and $m_B(\xi)$ oscillate around either ±1/2, that corresponds to the f phase, or zero values which corresponds to the p phase; hence we have the coexistence solution (f + p), as explained above. In Fig. 1(e), $m_A(\xi)$ oscillates around ±1/2 and $m_B(\xi)$ oscillates around ±3/2, which corresponds to the i phase, and also $m_A(\xi)$ and $m_B(\xi)$ are equal to each other and they oscillate around zero value, this solution corresponds to the p phase; hence we have the coexistence solution (i + p). A symmetric solution does not depend on the initial values, but the other solutions depend on the initial values. Finally we should also mention that the ferromagnetic phase has been defined as $m_A\neq m_B\neq 0$ in general [19], but in a few work, it was defined as $m_A\neq -m_B\neq 0$ [20].

In order to see the dynamic boundaries among these phases, we have to calculate DPT points and then we can present the phase diagrams of the system. DPT points will be obtained by investigating the behavior of the average magnetizations in a period or the dynamic magnetizations as a function of the reduced temperature. The dynamic order parameters, namely dynamic sublattice magnetizations ($M_A$, $M_B$) are defined as

$$M_A = \frac{1}{2\pi}\int_0^{2\pi} m_A(\xi)d\xi \qquad \text{and} \qquad M_B = \frac{1}{2\pi}\int_0^{2\pi} m_B(\xi)d\xi. \qquad (18)$$

The behaviors of $M_A$ and $M_B$ as a function of the reduced temperature for several values of d and h are obtained by combining the numerical methods of Adams-Moulton predictor corrector with the Romberg integration. A few interesting results are plotted in Figs. 2(a)-(d) in order to illustrate the calculation of the DPT points and the dynamic phase boundaries among the phases. $T_C$ and $T_{C'}$ are the second-order phase transition temperature from the i phase to the p phase, and from the f phase to the p phase, respectively. $T_t$ represents the first-order phase transition temperature. Fig. 2(a) shows the behavior of $M_A$ and $M_B$ as a function of the reduced temperature for d = 0.125 and h = 0.125. In this figure, $M_A = 1/2$ and $M_B = 3/2$ at zero temperature, and they decrease to zero continuously as the reduced temperature increases, therefore a second-order phase transition occurs at $T_C = 0.555$. In this case the dynamic phase transition is from the i phase ($M_A \neq M_B \neq 0$) to the p phase ($M_A = M_B = 0$) and the solution does not depend on initial values of $M_A$ and $M_B$. Fig. 2(b) presents the thermal variations of $M_A$ and $M_B$ for d = -0.5 and h = 0.125. In Fig. 2(b), $M_A = M_B = 1/2$ at zero temperature, and they decrease to zero continuously as the reduced temperature increases, therefore a second-order phase transition occurs at $T_{C'} = 0.265$ from the f phase to the p phase. This solution does not also depend on initial values of $M_A$ and $M_B$. Figs. 2(c) and (d) illustrate the thermal variations of $M_A$ and $M_B$ for d = 0.125 and h = 0.575 for two different initial values; i.e., the initial values of $m_A = 1/2$ and $m_B = 3/2$ for Fig. 2(c), and $m_A = m_B = 1/2$ or zero for Fig. 2(d). The behavior of Fig. 2(c) is similar to Fig. 2(a), hence the system undergoes a second-order phase transition from the i phase to the p phase at $T_C = 0.2875$. In Fig. 2(d), $M_A = M_B = 0$ at zero temperature, the system undergoes two successive phase transition as the temperature increases: The first one is a first-order phase transition, because discontinuity occurs for the dynamic magnetizations, and the transition is from the p phase to the i phase at $T_t = 0.2125$. The second one is a second-order phase transition from the i phase to the p phase at $T_C = 0.2875$ as similar to Figs. 2(a) and (c). From Figs. 2(c) and (d), one can see that the i + p coexistence region also exists in the system and this fact is seen in the phase diagram of Fig. 5(a), explicitly.

It is worth mentioning that if the single Ising [21] or mixed Ising [22] systems are in the static magnetic field, the systems do not undergo any phase transition within the mean-field approach. This fact is also correct for our calculation in this work that has been shown in our previous paper of the single spin-1 Blume-Capel (BC) model [23]. Now, we have also checked this fact for the mixed spin-1/2 and spin-3/2 Ising ferrimagnetic model, namely we have investigated the behavior of the dynamic order parameters in the static external magnetic field. Fig. 3 shows the thermal variations of $M_A$ and $M_B$ for several values of static h and d = − 0.125; hence this figure indicates that the system does not undergo any phase transition. These behaviors are similar to Fig. 6 (a) of Ref. 23, compare Fig. 3 with Fig. 6 (a) of Ref. 23.

The behaviors of dynamic magnetizations as a function of the reduced crystal-field interaction or single-ion anisotropy (d) are also investigated and presented four representative graphs, seen in Fig. 4. Fig. 4(a) is obtained for h = 0.375 and T = 0.25, and the system undergoes a second-order phase transition at $d_C = − 0.3825$, because $M_A$ and $M_B$ become zero continuously. Figs. 4(b) and (c) are calculated for h = 0.625 and T = 0.1 for two different initial values; i.e., the initial values of $m_A = 1/2$ and $m_B = 3/2$ for Fig. 4(b) and $m_A = m_B = 1/2$ or zero for Fig. 4(c). In Fig. 4(b), the system undergoes two successive phase transitions; the first one is a first-order phase transition and the transition is from the p phase to the i phase at $d_{t1} = 0.00$, and the second one is a second-order phase transition from the i phase to the p phase at $d_C = -0.285$. The behavior of Fig. 4(c) is similar to Fig. 4(b), but the first-order phase transition occurs at $d_{t2} = -0.2075$. From Figs. 4(b) and 4(c) one can see that the p phase until $d_{t1} = 0.00$; the i + p coexistence phase between $d_{t1} = 0.00$ and $d_{t2} = -0.2075$; the i phase between $d_{t2} = -0.2075$ and $d_C = -0.285$; after $d_{t2} = -0.2075$ the p phase, exist in the system and this fact is seen in the phase diagram of Fig. 6(c) explicitly [compare in Figs. 4(b) and 4(c) with Fig. 6(c)]. Fig. 4(d) displays the behaviors of magnetizations for $h = 0.125$ and $T = 0.05$. At the high values of a reduced crystal-field interaction, $M_A = 1/2$ and $M_B = 3/2$; hence we have the ferrimagnetic (i) phase, and as the reduced crystal-field decreases the i phase becomes the ferromagnetic (f) phase ($M_A = M_B = 1/2$) with the second-order phase transition $d_{C'} = -0.3125$.

**4. Dynamic phase diagrams**

Since we have obtained the DPT points in Section 3, we can now present the phase diagrams of the system. The calculated phase diagrams in the (h, T) and (d, T) planes are presented in Figs. 5 and 6, respectively for various values of interaction parameters. In these phase diagrams, the solid and dashed lines represent the second- and first-order phase transition lines, respectively, and the dynamic tricritical points are also denoted by a solid circle. The dotted line is an ordered line smoothly mediating, with no phase transition, between the different ordered phases.

In Fig. 5, only one dynamic tricritical point exists and two different topological types of phase diagrams are found. **(i)** Fig. 5(a) represents the phase diagram in the (h, T) plane for d = 0.125. In this phase diagram, at high reduced temperature (T) and high reduced external magnetic field (h), the solutions are paramagnetic (p); and at low values of T and h, are ferrimagnetic (i). The dynamic phase boundary between these regions, i → p, is the second-order phase transition line. At low reduced temperatures, there is a range of values of h in which the p and i phases or regions coexist, called the coexistence or mixed region, i + p. The i + p region is separated from the i and the p phases by the first-order phase transition lines. The system also exhibits only one dynamic tricritical point where the both first-order phase transition lines merge and signals the change from the first- to the second-order phase transition. Finally, we should also mention that very similar phase diagrams were also obtained in kinetics of the mixed spin-1/2 and spin-1 Ising ferrimagnetic system [24], the kinetic spin-1 Ising systems [23, 25] and the kinetic spin-3/2 Ising systems [26], but the phases other than the p phases are different. **(ii)** Fig. 5 (b) calculated for d = - 0.5 and it is similar to Fig. 5(a), except that the i + p phase becomes f + p phase and the i phase turns to the f phase.

The calculated phase diagrams of the system in the (d, T) are seen in Figs. 6 (a)-(c). As seen in Fig.6, we have obtained three different phase diagram topologies. **(i)** For h = 0.125, we are performed the phase diagram, seen in Fig. 6(a). The system always undergoes a second-order phase transition. Besides one dynamic multicritical point (A), the p, f and i phases exist in the phase diagram. The dynamic phase boundaries among the p, f and i are the second-order phase transition lines. For high values of T, the p phase always exists, but low values of T and large negative values of d, the f phase exists and for low values of T and high values of d, the i phase occurs. We have found a similar dynamic phase diagram to the one obtained in the kinetic spin-3/2 BC model [27], except the following differences: (1) The i phase becomes the $f_{3/2}$ phase, (2)

For very low values of T and d, the $f_{3/2} + f_{1/2}$ coexistence phase exists and the dynamic phase boundary between the $f_{3/2} + f_{1/2}$ and $f_{3/2}$, and between the $f_{3/2} + f_{1/2}$ and $f_{1/2}$ phase are first-order phase lines. Moreover, we have also found the similar phase diagram, except the second-order phase transition line between the f and i phases becomes a first-order line, to the one obtained by methods in the equilibrium statistical physics in spin-3/2 Ising systems, namely the mean-field approximation and the Monte Carlo simulation [28], a renormalization-group transformation in position-space based on the Migdal-Kadanoff recursion relations [29], the cluster expansion method [30] and in the exact solution of the model on the Bethe lattice by using the exact recursion equations [31]. **(ii)** For h = 0.375, the phase diagram is constructed in Fig. 6(b) and is similar to the phase diagram of Fig. 6(a) but following differences have been found: (1) The second-order phase line and the f phase occur at low temperatures disappear. (2) Two more coexistence phases, namely the f + p, i + p phases, occur for very low values of T, and the dynamic phase boundary between these two mixed phases is a second-order line. (3) The dynamic phase boundaries between the f + p and p phases, and between the i + p and the i phases are the first-order phase lines. (4) The dynamic critical end point (E) appears instead of the dynamic multicritical point (A). (5) The dynamic tricritical points, where the both first-order phase transition lines merge and signals the change from the first- to the second-order phase transitions, occurs. **(iii)** For h = 0.625, the phase diagram is given in Fig. 6(c). This phase diagram exhibits the p, i and i + p phases besides the two dynamic tricritical points. The dynamic phase boundary between the i and p phase is a second-order line that occurs for negative values of d, and all other phase lines among the other phases are first-order lines.

**5. Summary and Conclusion**

We have analyzed, within a mean-field approach, the stationary states of the kinetic mixed spin-1/2 and spin-3/2 Ising ferrimagnetic model with a crystal-field interaction under the presence of a time varying (sinusoidal) magnetic field. We use a Glauber-type stochastic dynamics to describe the time evolution of the system. First we have studied time variations of the average magnetizations in order to find the phases in the system. Then, the behavior of the dynamic magnetizations as a function of the reduced temperature and a crystal-field interaction is investigated to find the nature of phase transitions and as well as to calculate DPT points. The dynamic phase diagrams are presented in the (h, T) and (d, T) planes. We have found that the behavior of the system strongly depends on the values of the interaction parameters and two different phase diagram topologies are obtained in the (h, T) plane and three fundamental phase diagrams are found in the (d, T) plane. The phase diagrams exhibit the p, f, i, f+p and/or i+p coexistence regions depending on the interaction parameter values and the dynamic phase boundaries among these phases are first-order lines for most cases and second-order lines for a few cases. Therefore, the phase diagrams always exhibits dynamic tricritical points in the (h, T) plane, but does not exhibit in the (d, T) plane for low values of h, seen in Fig. 6(a). Moreover, the dynamic critical end point (E) and dynamic multicritical point (A) exist in the (d, T) plane for low values of h, seen in Fig. 6 (a) and (b), respectively.

Finally, it should be mentioned that this mean-field dynamic study, in spite of its limitations such as the correlation of spin fluctuations have not been considered, suggests that the kinetic mixed spin-1/2 and spin-3/2 Ising ferrimagnetic model with crystal field has an interesting dynamic behavior. Hence, we hope that our detailed theoretical investigation may stimulate further works to study the nonequilibrium or the dynamic phase transition (DPT) in the mixed Ising model by using the dynamic Monte Carlo (MC) simulations in which our results will be instructive for the time consuming process searching critical behavior of this system while using the dynamic MC simulations. We also mention that some of the first-order lines and as well as tricritical points might be artifact of the mean-field calculation, this fact has been discussed extensively in the kinetic spin-1/2 Ising model in the recent works [32-34]; hence this system should be studied by non-perturbative methods, such as MC simulations and renormalization-group (RG) calculations in order to find the artifact first-order phase line as well as the tricritical point.

**Acknowledgements**

This work was supported by the Scientific and Technological Research Council of Turkey (TÜBİTAK), Grant No: 107T533 and Erciyes University Research Funds, Grant No: FBA-06-01. One of us (B.D.) would like to express his gratitude to the TÜBİTAK for the Ph.D scholarship.

### List of the Figure Captions

**Fig. 1.** Time variations of the average magnetizations ($m_A$, $m_B$):

a) Exhibiting a paramagnetic (p) phase: $d = -0.5$, $h = 0.25$ and $T = 0.375$.
b) Exhibiting a ferromagnetic-1/2 (f) phase: $d = -0.5$, $h = 0.15$ and $T = 0.10$.
c) Exhibiting a ferrimagnetic (i) phase: $d = 0.125$, $h = 0.20$ and $T = 0.50$.
d) Exhibiting a coexistence region (f+p): $d = -0.5$, $h = 0.40$ and $T = 0.05$.
e) Exhibiting a coexistence region (i+p): $d = 0.125$, $h = 0.60$ and $T = 0.025$.

**Fig. 2.** The reduced temperature dependence of the dynamic magnetizations, $M_A$ and $M_B$. The $T_C$ is the second-order phase transition temperature from the i phase to the p phase; $T_{C'}$ is from the f phase to the p phase; $T_t$ represents the first-order phase transition temperature from the i phase to the p phase.

**a)** Exhibiting a second-order phase transition from the i phase to the p phase for $d = 0.125$ and $h = 0.125$; 0.555 is found $T_C$.

**b)** Exhibiting a second-order phase transition from the f phase to the p phase for $d = -0.5$ and $h = 0.125$; 0.265 is found $T_{C'}$.

**c)** Exhibiting a second-order phase transition from the i phase to the p phase for $d = 0.125$, $h = 0.575$ and the initial values of $M_A = 1/2$ and $M_B = 3/2$; 0.2875 is found $T_C$.

**d)** Exhibiting two successive phase transition, the first one is a first-order phase phase transition from the p phase to the i phase and the second one is a second-order phase transition from the i phase to the p phase for $d = 0.125$, $h = 0.575$ and the initial values of $M_A = M_B = 1/2$ or zero; 0.2125 and 0.2875 are found $T_t$ and $T_C$, respectively.

**Fig. 3.** Thermal variations of the dynamic order parameters for several values of the static external magnetic fields h and $d = -0.125$.

**Fig. 4.** The behavior of dynamic magnetizations as a function of the reduced crystal-field interaction or single-ion anisotropy.

   a) Exhibiting a second-order phase transition from the i phase to the p phase for h = 0.375 and T = 0.25; - 0.3825 is found $d_C$.

   b) Exhibiting two successive phase transitions, the first one is a first-order phase transition from the p phase to the i phase and the second one is a second-order phase transition from the i phase to the p phase for h = 0.625 and T = 0.1 and the initial values of $M_A$ = 1/2 and $M_B$ = 3/2 ; 0.00 and - 0.285 are found $d_{t1}$ and $d_C$, respectively.

   c) Same as (b) but the initial values of $M_A$ = $M_B$ = 1/2 or zero; - 0.2075 and - 0.285 are found $d_{t2}$ and $d_C$, respectively.

   d) Exhibiting a second-order phase transition from the f phase to the i phase for h = 0.125 and T = 0.05; - 0.3125 is found $d_{C'}$.

**Fig. 5.** Phase diagrams of the mixed spin-1/2 and spin-3/2 Ising ferrimagnetic model in the (h, T) plane. The paramagnetic (p), ferromagnetic (f), ferrimagnetic (i) and two different coexistence or mixed phases, namely the i+p and f+p phases, are found. Dashed and solid lines represent the first- and second-order phase transitions, respectively, and dynamic tricritical point is indicated with a filled circle. **a)** d = 0.125, **b)** d = - 0.50.

**Fig. 6.** Same as Fig. 5, but in the (d, T) plane. **a)** h = 0.125, **b)** h = 0.375, **c)** h = 0.625.

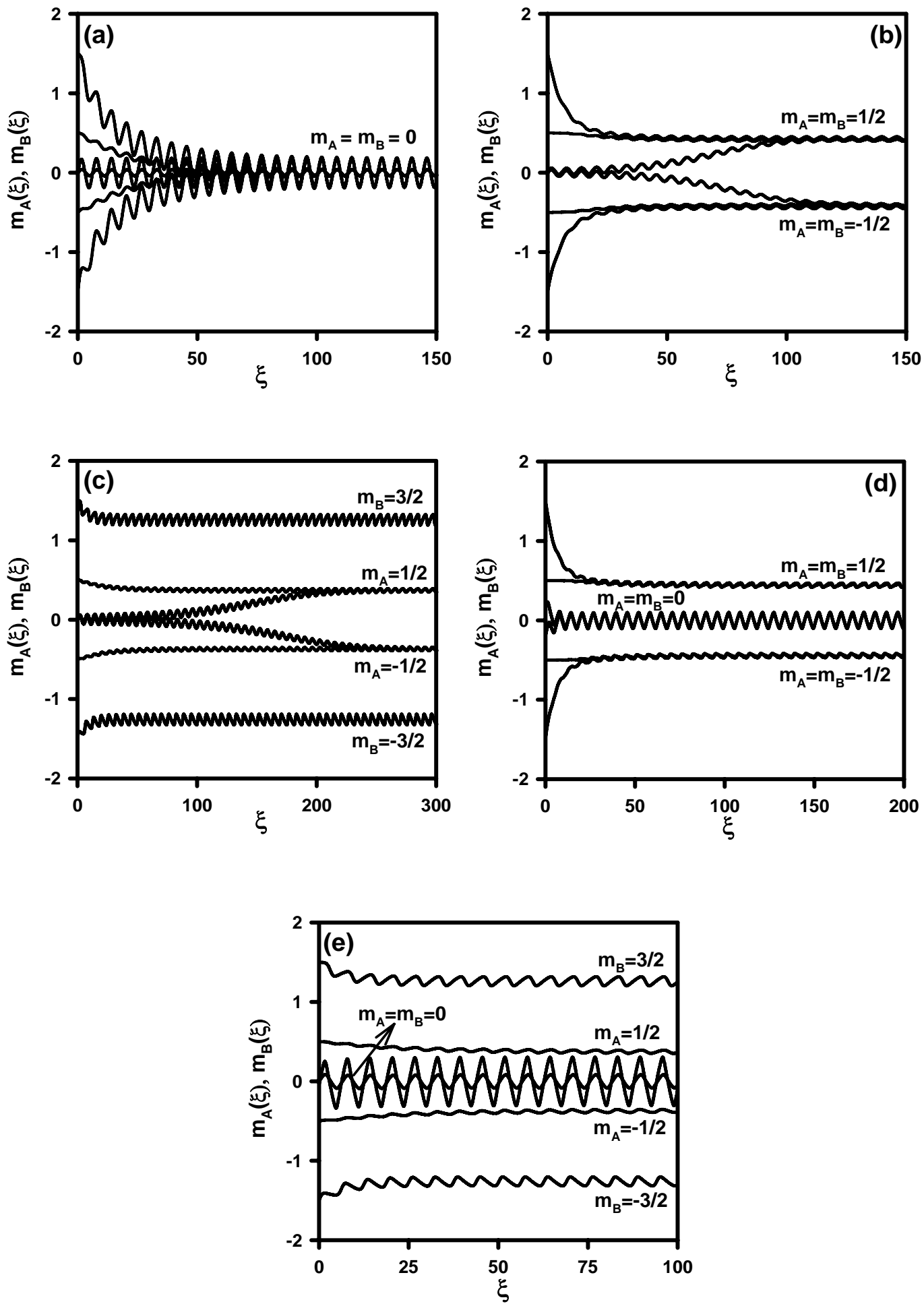

Fig. 1

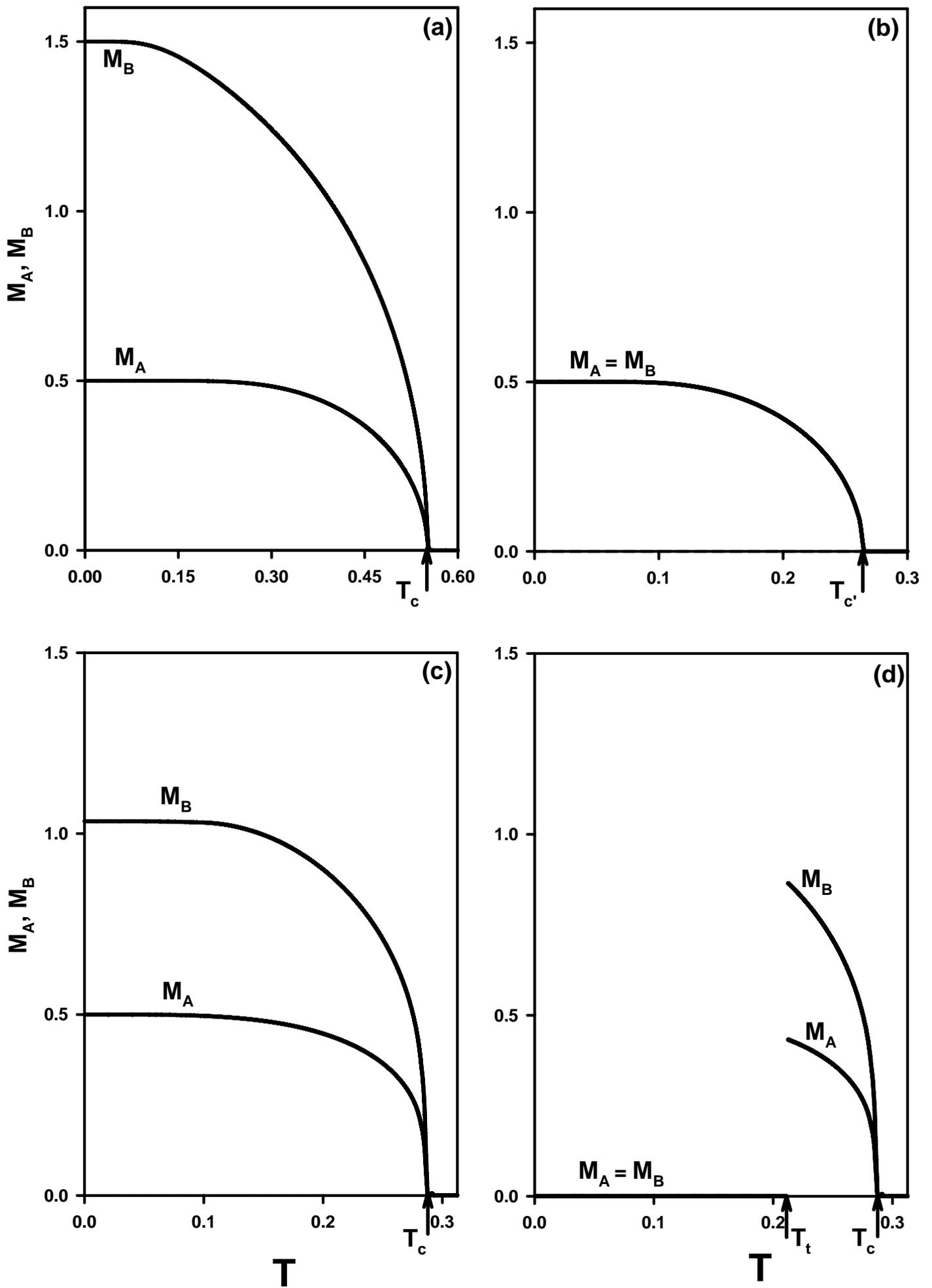

Fig. 2

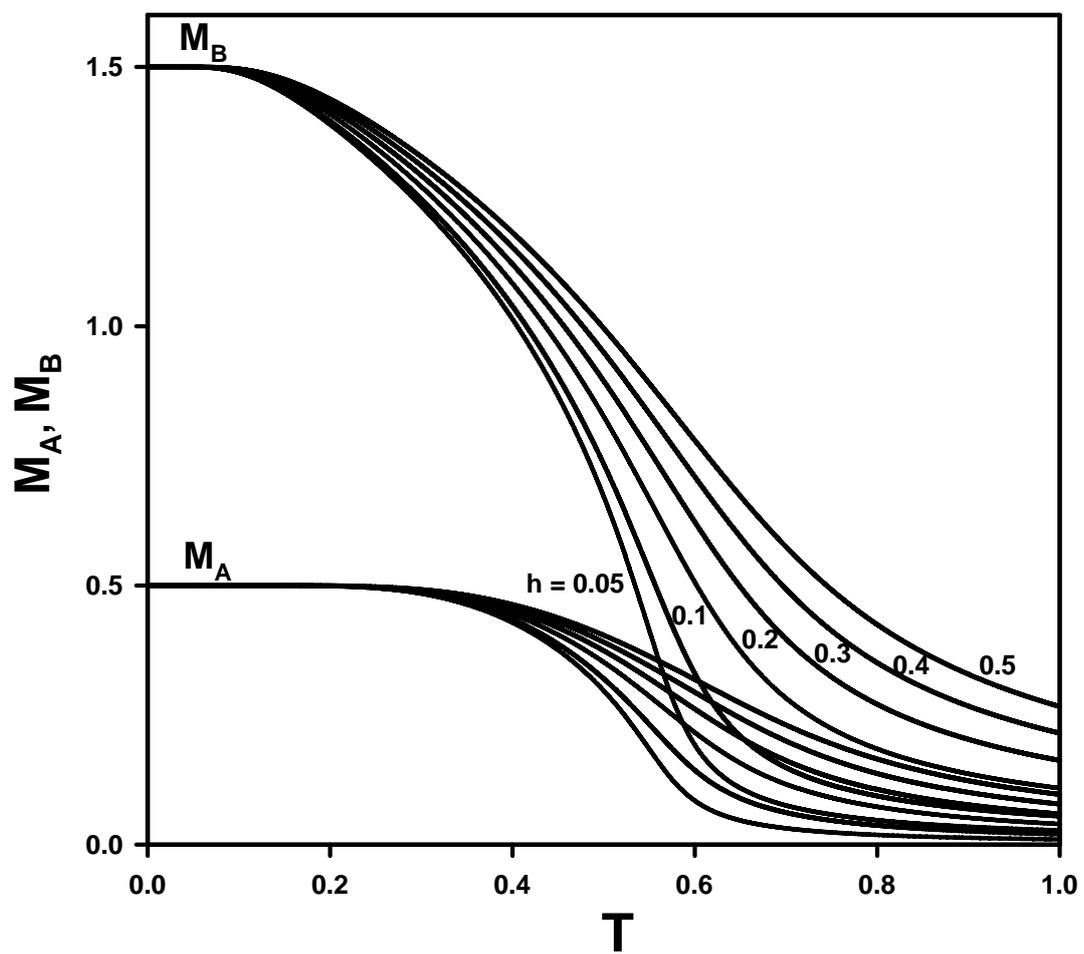

Fig. 3

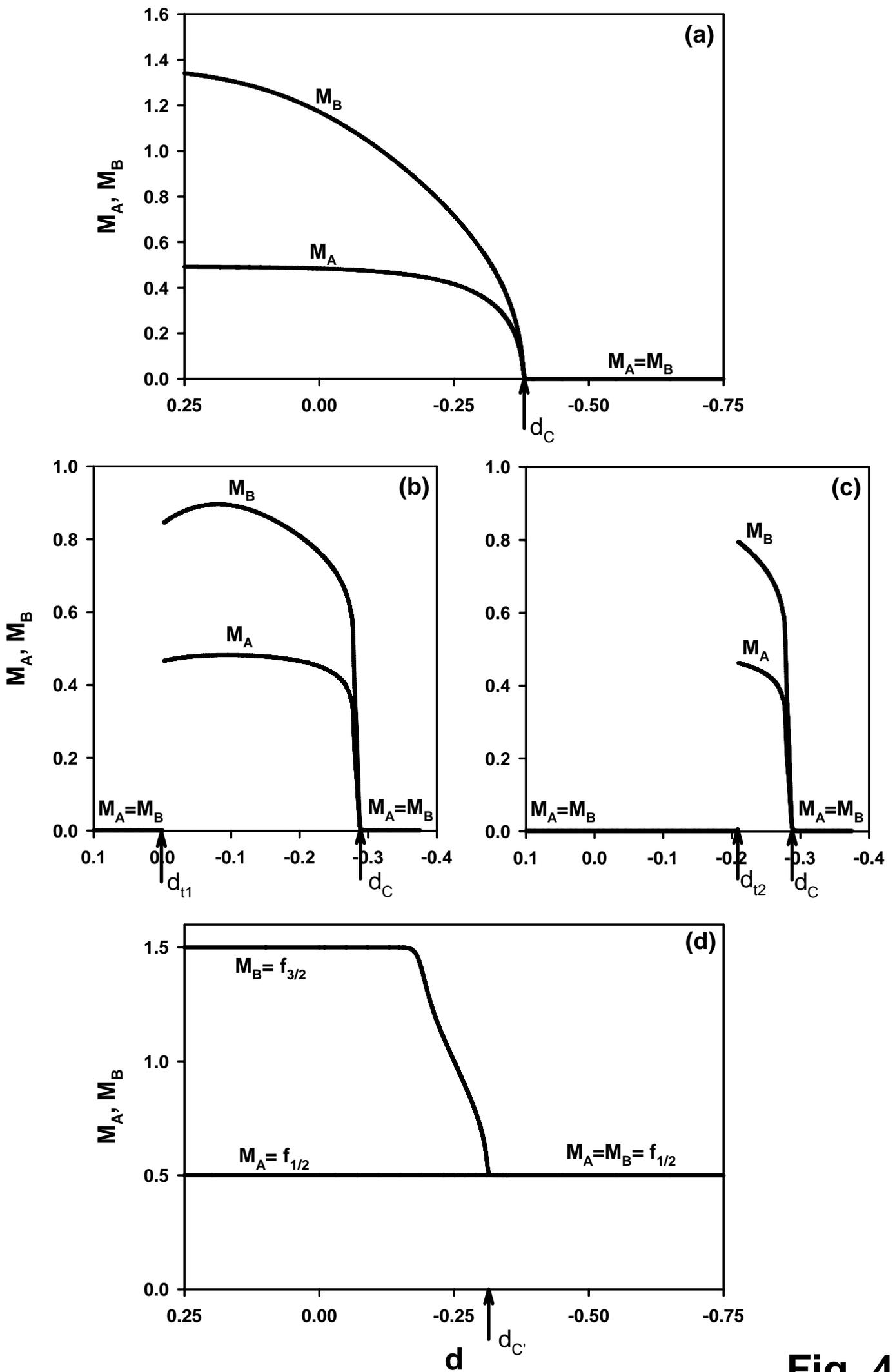

Fig. 4

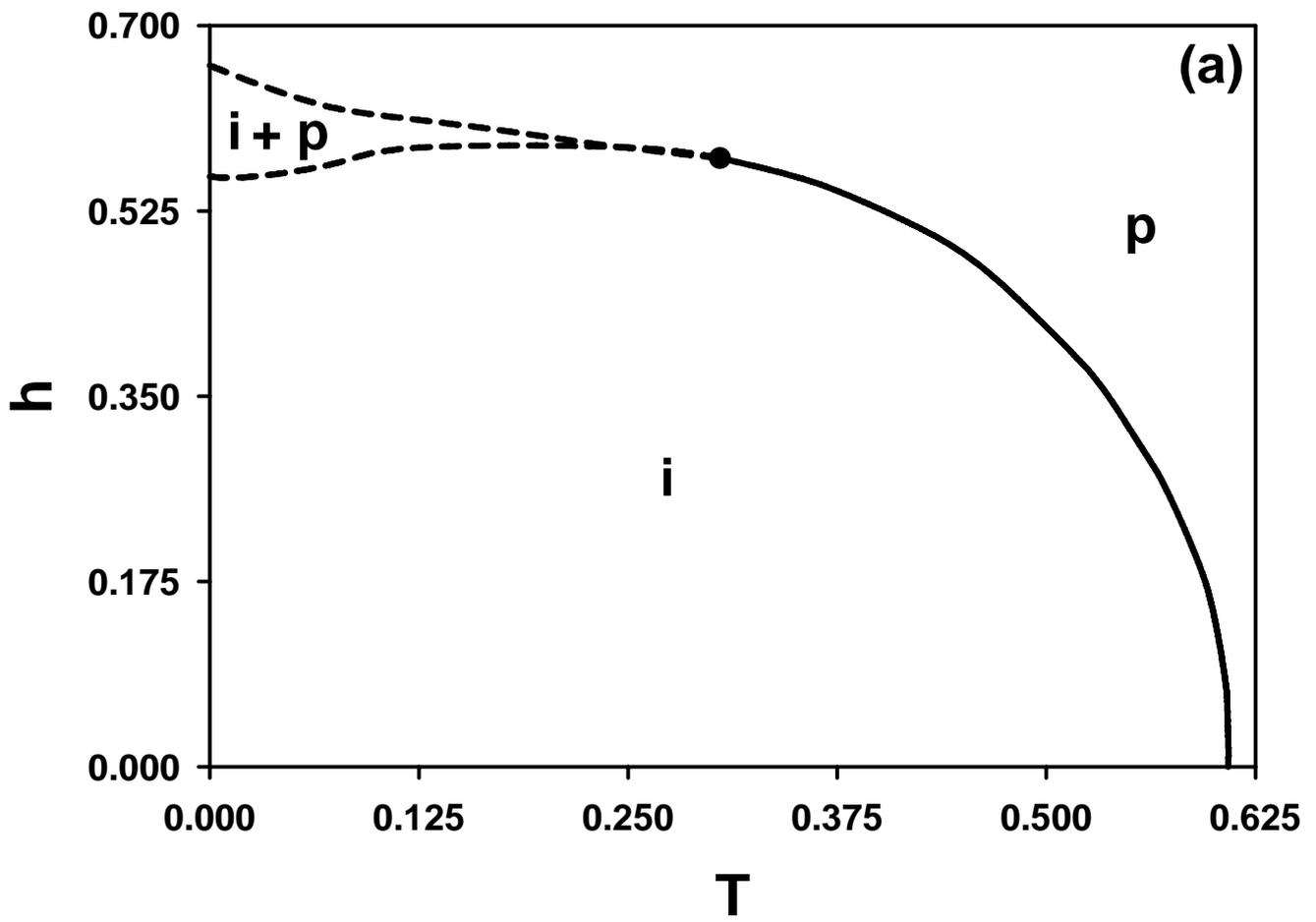

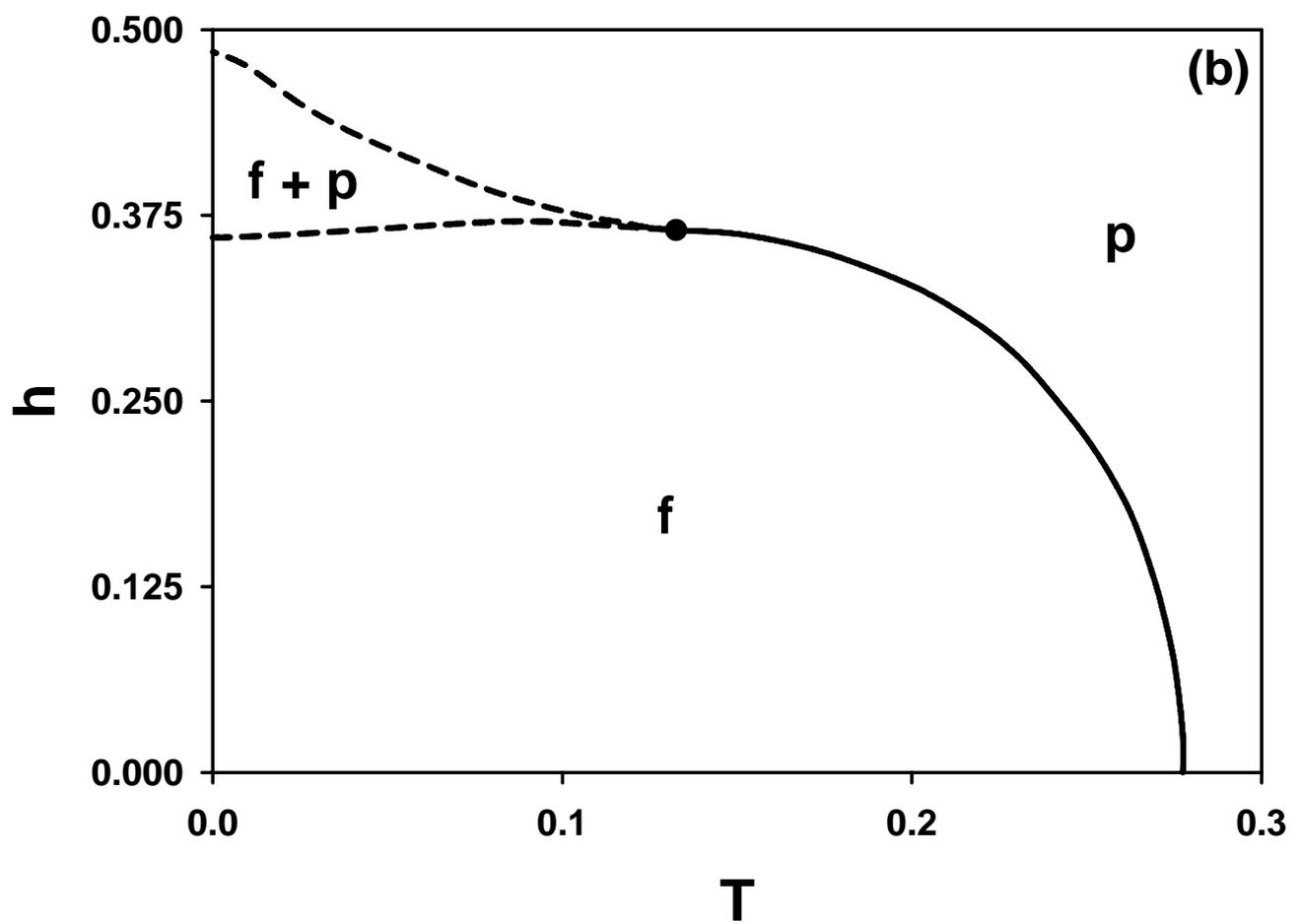

Fig. 5

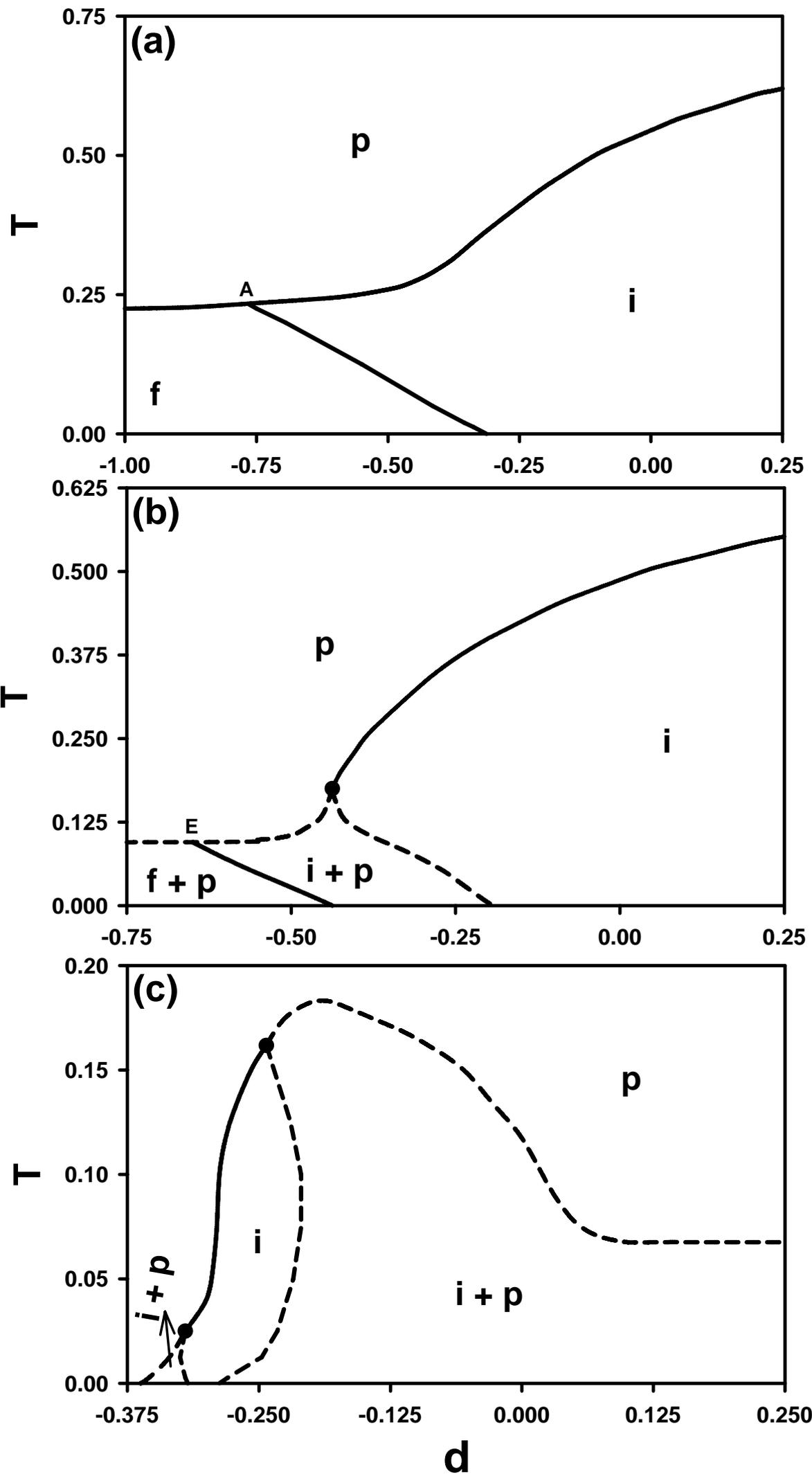

Fig. 6